\documentclass[12pt,preprint,twocolumn]{aastex6}
\usepackage[utf8]{inputenc}
\usepackage{soul}
\usepackage{xcolor}

\usepackage[normalem]{ulem}

\begin{document}
\title{PGC 38025: A star-forming lenticular galaxy with an off-nuclear star-forming core}

\author{Zhengyi Chen\altaffilmark{1, 2}, Qiu-Sheng Gu $^{\color{blue} \dagger}$\altaffilmark{1, 2},  Rub\'en Garc\'ia-Benito\altaffilmark{3}, Zhi-Yu Zhang\altaffilmark{1, 2}, Xue Ge\altaffilmark{4}, Mengyuan Xiao\altaffilmark{1, 2, 5}, Xiaoling Yu\altaffilmark{1, 2}}
\altaffiltext{1}{School of Astronomy and Space Science, Nanjing University, Nanjing, Jiangsu 210093, China (Email: qsgu@nju.edu.cn)}
\altaffiltext{2}{Key Laboratory of Modern Astronomy and Astrophysics (Nanjing University), Ministry of Education, Nanjing 210093, China}
\altaffiltext{3}{Instituto de Astrof\'isica de Andaluc\'ia (CSIC), P.O. Box 3004, 18080 Granada, Spain}
\altaffiltext{4}{School of Physics and Electronic Engineering, Jiangsu Second Normal University, Nanjing, Jiangsu 211200, China}
\altaffiltext{5}{AIM, CEA, CNRS, Universit\'e Paris-Saclay, Universit\'e Paris Diderot, Sorbonne Paris Cit\'e, F-91191 Gif-sur-Yvette, France}

\date{Accepted ??. Received ??}

\begin{abstract}
Lenticular galaxies (S0s) were considered mainly as passive evolved spirals due to environmental effects for a long time; however, most S0s in the field cannot fit into this common scenario. In this work, we study one special case, SDSS J120237.07+642235.3 (PGC 38025), a star-forming field S0 galaxy with an off-nuclear blue core. We present optical integral field spectroscopic (IFS) observation with the 3.5 meter telescope at Calar Alto (CAHA) Observatory, and high-resolution millimeter observation with the NOrthern Extended Millimeter Array (NOEMA). We estimated the star formation rate (SFR = 0.446 $M_\odot yr^{-1}$) and gaseous metallicity (12 + log(O/H) = 8.42) for PGC 38025, which follows the star formation main sequence and stellar mass - metallicity relation. We found that the ionized gas and cold molecular gas in PGC 38025 show the same spatial distribution and kinematics, whilst rotating misaligned with stellar component. The off-nuclear blue core is locating at the same redshift as PGC 38025 and its optical spectrum suggest it is \rm H\,{\sc ii} region. We suggest that the star formation in PGC 38025 is triggered by a gas-rich minor merger, and the off-nuclear blue core might be a local star-formation happened during the accretion/merger process.
\end{abstract}


\section{Introduction} 
\label{sec:introduction}
Lenticular galaxies (S0s) are morphologically situated between ellipticals and spirals, e.g., armless and with dense cores (bulges) as ellipticals, while having disk component as spirals do~\citep{1936rene.book.....H, 1976ApJ...206..883V, 1961hag..book.....S}. They are traditionally considered as having used up most of their gas reservoir which is critical for star formation.~\citep{2007ApJ...671.1624D, 1993AJ....106..473C}.

S0s in a group and/or cluster (i.e., dense environment) are considered to be transformed from spirals due to environmental effects, e.g., ram-pressure stripping in hot intergalactic medium (IGM)~\citep{1972ApJ...176....1G, 2000Sci...288.1617Q}, gravitational tidal effect and harassment~\citep{1983ApJ...265..664D, 1990ApJ...350...89B, 1996Natur.379..613M, 2002ApJ...565..223S}, and encounter of galaxies~\citep{1951ApJ...113..413S, 1985A&A...144..115I}. S0s tend to be in the dense environment. \citet{1984ApJ...281...95P} found there are 40$\%$ - 50$\%$ S0s in group environment, and the fraction raises to 60$\%$ in clusters~\citep{1980ApJ...236..351D, 1984ApJ...281...95P}. However, only 15$\%$ nearby  field galaxies are S0s~\citep{1995MNRAS.274.1107N}. These relatively rare field S0s cannot be fitted into the common formation scenario (environmental triggered formation) of S0s in groups/clusters, and may contribute to the diversity in the properties of S0s. Recent galaxy merger, gas accretion into elliptical galaxies, secular evolution including instabilities and stellar feedback are proposed as possible formation scenarios of these field S0s instead~\citep{1998ApJ...502L.133B, 2005A&A...437...69B, 2017MNRAS.471.1892B, 2018A&A...617A.113E}.

Recent studies reveal something intriguing on S0s. $\sim$75$\%$ S0 galaxies are with CO detection in~\cite{2003ApJ...584..260W}. Cold gas, in atomic and/or molecular phase, may be presented in most of S0s~\citep{2006ApJ...644..850S, 2010ApJ...725..100W}. Rejuvenation or recent star formation triggered by gas accretion is found in the local S0 galaxies~\citep{2010ApJ...714L.171T, 2016ApJ...831...63X, 2019ApJS..244....6S, 2020ApJ...889..132G}, and the sources of cold gas accretion may be merged with gas-rich dwarf satellites~\citep{2009MNRAS.394.1713K, 2011MNRAS.411.2148K} and cosmological filaments~\citep{2006MNRAS.368....2D, 2005MNRAS.363....2K}. Many S0 galaxies hold rather complex stellar population, star formation history and kinematics, which make it inconceivable to treat S0 as simple transitional stage between elliptical and spiral galaxies~\citep{2018ApJ...862..100G}. 

Studying the evolution of S0 galaxies is essential to understanding formation and evolution of all different types of galaxies. Furthermore, the study of isolated star-forming S0 by using deep optical IFS data can provide valuable information not only for the central part but also for the disc component of a galaxy. This is an essential issue because it provides a crucial breakthrough point for testing formation scenarios for S0s.

\begin{figure}
    \centering
    \includegraphics[width = 1.0\linewidth]{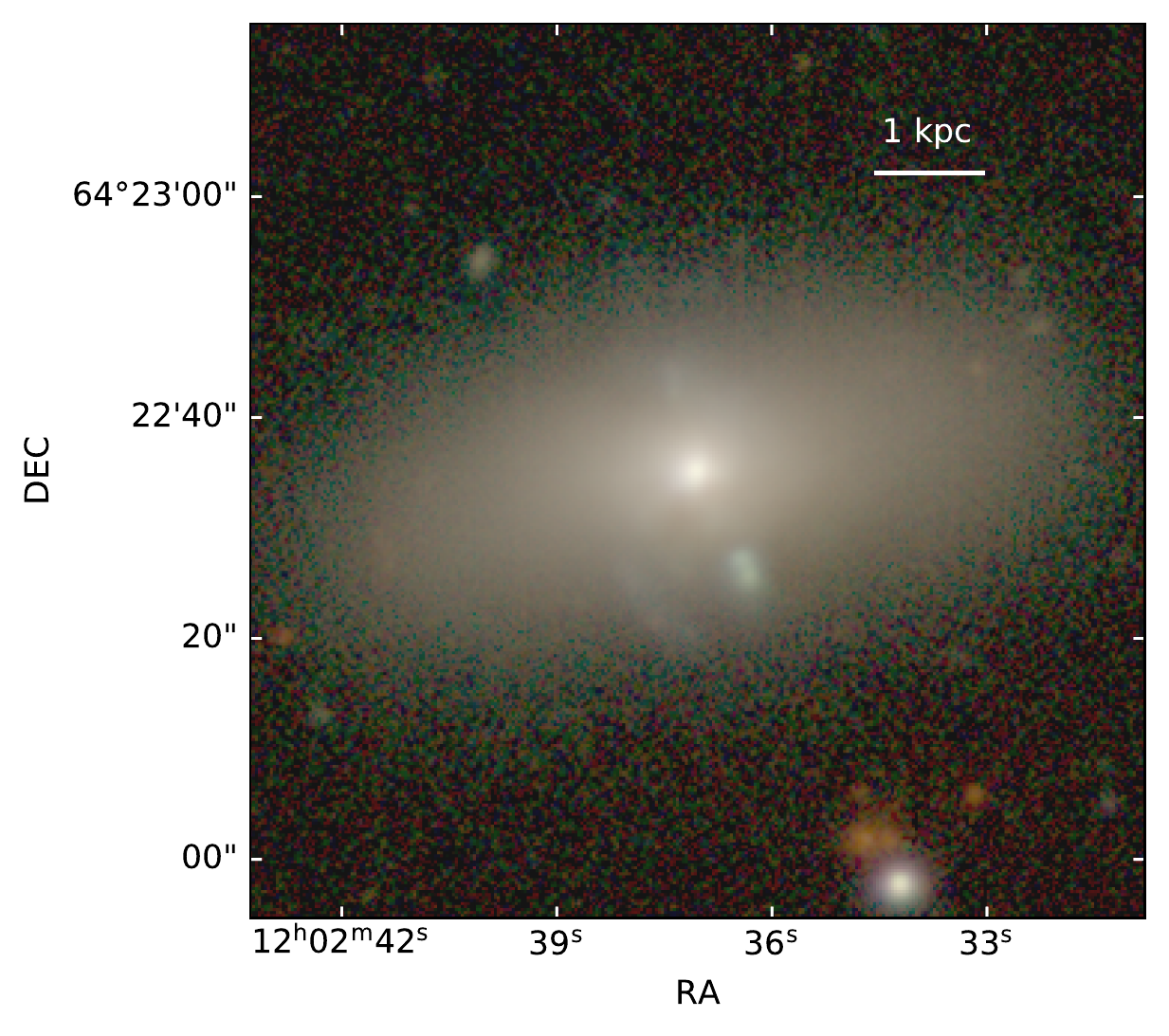}
    \caption{Optical grz composite image from DESI Legacy Imaging Surveys~\citep{2019AJ....157..168D} }
    \label{fig:Legacy survey}
\end{figure}

SDSS J120237.07+642235.3 (PGC 38025) is a star-forming field lenticular galaxy with an off-nuclear blue core ($\sim$ 1 kpc away from the galaxy centre), as shown in Figure~\ref{fig:Legacy survey}. \citet{2016ApJ...831...63X} revealed central star formation in PGC 38025. Its early-type morphology, together with the extended star-formation and off-nuclear blue core make PGC 38025 to be an ideal target in studying formation and evolution of S0s. The CAHA optical IFS observation we obtained, compared to single aperture spectroscopy such as Sloan Digital Sky Survey (SDSS), is highly competent in the study on spatial extended star-formation mechanism of PGC 38025 and nature of its off-nuclear blue core. In order to study the origin of the star forming activity in  this galaxy, we also carried out the interferometer radio observation from IRAM NOEMA to reveal molecular gas content and spatial distribution of PGC 38025.

This paper is organised as follows, Section~\ref{sec:observation} presents CAHA optical and IRAM-NOEMA Observation. Section~\ref{sec:Results} gives main results of observational data analysis. Discussion and summary are in Section~\ref{sec: discussion} and \ref{sec: summary}. We assume a flat $\rm\Lambda$CDM cosmology with $\rm{\Omega}_{m}$ = 0.3, $\rm{\Omega}_{\Lambda}$ = 0.7, and $\rm{H}_{0}$ = 70 $\rm{km}^{-1}$ $\rm{s}^{-1}$ $\rm{Mpc}^{-1}$ in this paper~\citep{2013ApJS..208...19H}.

\section{Observational Data}
\label{sec:observation}
\def\th{\tablenotemark{h}}
\begin{deluxetable}{ll}
    \setlength{\tabcolsep}{4pt}
    \tablewidth{0in}
    \tablecolumns{11}
    \tablecaption{Description of NOEMA observations\label{tab:obslog}}
    \tablehead{
    \multicolumn2c{PGC 38025}
    }
    \startdata
    RA  	                    & 12: 02: 37.195 \\
    Dec 			        	& 64: 22: 29.070 \\
    Redshift                	& 0.00505 \\
    Obs. date					& 2019 July 2, 4, 8  \\
                                & September 9, 14\\
    Configuration				& D \\
    N$_{ant}$					& 9 \\
    Obs. freq (GHz) 			& 114.692 \\
    Time on source (hr)			& 6.8 \\
    FoV (arcsec)				& 43.9$\times$43.9 \\
    Synth. beam (arcsec)		& 3.86$\times$2.96  \\
    Synth. beam (kpc)			& 0.40$\times$0.31  \\
    \enddata
\end{deluxetable}

\subsection{CAHA Optical IFU Spectroscopic Observation}
Optical observations of PGC 38025 were performed using PPAK Integral Field Unit (IFU) of Potsdam Multi-Aperture Spectrograph (PMAS) mounted on the 3.5m telescope at the Calar Alto Observatory (CAHA). The observations were carried out on 2016 March 11 and 13 with two overlapping setups, i) low-resolution setup (V500; R$\sim$850) covering the wavelength range of 3745-7500 \AA, and ii) blue medium-resolution setup (V1200; R$\sim$1650) covering 3600-4620 \AA. A three-pointing dithering scheme was applied to both setups in order to reach a spatial filling factor of 100$\%$, and the exposure time per pointing was fixed to 900 s for the V500 and 2000 s (split in 2 individual exposures of 1000 s) for the V1200. The combination of spectroscopic cubes from two setups (called COMBO) was generated to reduce the interval vignetting effect on spectral edges within spectrograph. The final optical wavelength range of COMBO covers from 3700 to 7300 \AA.

Data reduction followed the common sequence of processes for fiber-fed spectrographs, 1) determination of spectra location on the CCD detector, 2) subtraction of scattered light and individual spectrum extraction, 3) dispersion correction to spectra, 4) correction for the fiber to fiber transition, 5) flux calibration and sky emission subtraction, 6) reconstruction of spectra in order of their original location in the sky, 7) atmospheric refraction and Galactic extinction correction. Data reduction was performed using a customized version of the pipeline of Calar Alto Legacy Integral Field spectroscopy Area survey (CALIFA). More details of the reduction process can be found in \citet{2013A&A...549A..87H},~\citet{2015A&A...576A.135G}, and~\citet{2016A&A...594A..36S}.

\subsection{IRAM-NOEMA Observation} 

We observed the CO (1-0) of PGC 38025 with the IRAM NOEMA interferometer on 2019 July 2nd, 4th and 8th, and September 9th and 14th(Project S19BL. PI: Zhengyi Chen). The observations were made with a total on source time of 6.8 hours in the D configuration in five tracks, which is the most compact one and with the maximum sensitivity (resolution $\sim$3.7'' at 100GHz and $\sim$1.6'' at 230GHz, suitable for detection experiments and coarse mapping). Further technical details of the observation are listed in Table~\ref{tab:obslog}. The tuning frequency was set on 114.692 GHz, the expected frequency of the redshift CO (1-0) line (\rm${\nu}_{rest}=115.271$ GHz) given a redshift of z = 0.00505 based on the optical observation. In addition, the flux calibrator MWC349 (tracks on July 2, 4, 8 and September 9) and LKHA101 (track on September 14) were observed. Bandpass calibrations are based on 3C454.3 (track on July 2), 3C345 (tracks on July 4, 8, and September 9), and 3C84 (track on September 14). Two phase calibrators, J1302+690 and 1030+611, are used for each track in order to guarantee robust spatial analysis.

The data reduction was performed using Continuum and Line Interferometer Calibration (CLIC; for calibration) and MAPPING (for imaging and deconvolution), which are modules of the Grenoble Image and Line Data Analysis Software (GILDAS\footnote{{\tt http://www.iram.fr/IRAMFR/GILDAS}}). Calibrations mainly include bandpass calibration, phase calibration, and flux calibration, and calibrators are mentioned above. The synthesized beam size is 3\farcs86 $\times$ 2\farcs96 at a position angle of 93$\deg$. We delivered lmv spectra cube (2 axes of coordinates and 1 axis of velocity/frequency) based on uv table, output of CLIC. We generated a 128 $\times$ 128 pixel map with 0\farcs62 per pixels, as recommended by GILDAS software based on the synthesized beam size. Frequency interval was re-sampled to be 2 MHz ($\sim$ 5.23 km $s^{-1}$).

\section{Results}
\label{sec:Results}
\subsection{Morphological Properties}
\label{Sec:Morphological Analysis}
We analyzed the surface brightness distribution of PGC 38025 utilizing the GALFIT \citep[version 3.0.5;][]{2002AJ....124..266P, 2010AJ....139.2097P}, and obtained the photometric parameters of the morphological components. GALFIT offers various two-dimensional models, e.g., the ``Nuker'' law, the S\'ersic (de~Vaucouleurs) bulge, and an exponential disk. Accurate model choices are essential for obtaining a good result. \citet{1948AnAp...11..247D} concluded that many elliptical galaxies have $R^{1/4}$ light distribution. In GALFIT this can be modeled with a S\'ersic profile represented by equation,
\begin{eqnarray}
I(R) = I_{\rm e}\  {\rm exp}\left\{-b_{\rm n}[(R/R_{\rm e})^{1/n} -1]\right\},
\label{sersic-eq}
\end{eqnarray}
\noindent
where R is the radial distance from the galaxy centre, I(R) is the surface brightness at radius R, $I_e$ is the surface brightness at half-light radius (also effective radius) $R_e$, n is the power-law index known as S\'ersic index (n = 4 in de Vaucouleur laws, and n = 1 for exponential disk) which variable $b_n$ is coupled with,
\begin{eqnarray}
b_{n} \simeq 1.9992n - 0.3271.
\label{bn1}
\end{eqnarray}
\noindent
\citet{1970ApJ...160..811F} stated that late-type galaxies (aside from elliptical galaxies) are composed of a de Vaucouleurs (S\'ersic index = 4) bulge and an exponential disk. As PGC 38025 is classified as a lenticular galaxy~\citep{2015ApJS..217...27A}, we adopt a classical bulge plus an exponential disk to fit.

There are some input files required in GALFIT, e.g., the observed image, the point spread function (PSF) of the image, a bad pixel mask, a background noise image, and initial guesses for the fitting parameters. We obtained SDSS g- and r-band observed and PSF image from NASA-Sloan Atlas\footnote{\tt http://www.nsatlas.org/data}. The background noise image can be automatically generated by the standard keywords, GAIN and NCOMBINE, in the header of observed image FITS file. 

We model the surface brightness of PGC 38025 with a bulge and a disk component, with initial  S\'ersic indexes for bulge (n = 4) and exponential disk (n = 1), and set them free to vary during GALFIT fitting iteration. A uniform sky pedestal is added, a value estimated utilizing SAOImage DS9~\citep{2003ASPC..295..489J}.

\begin{figure*}
	\centering
	\includegraphics[width=1.0\linewidth]{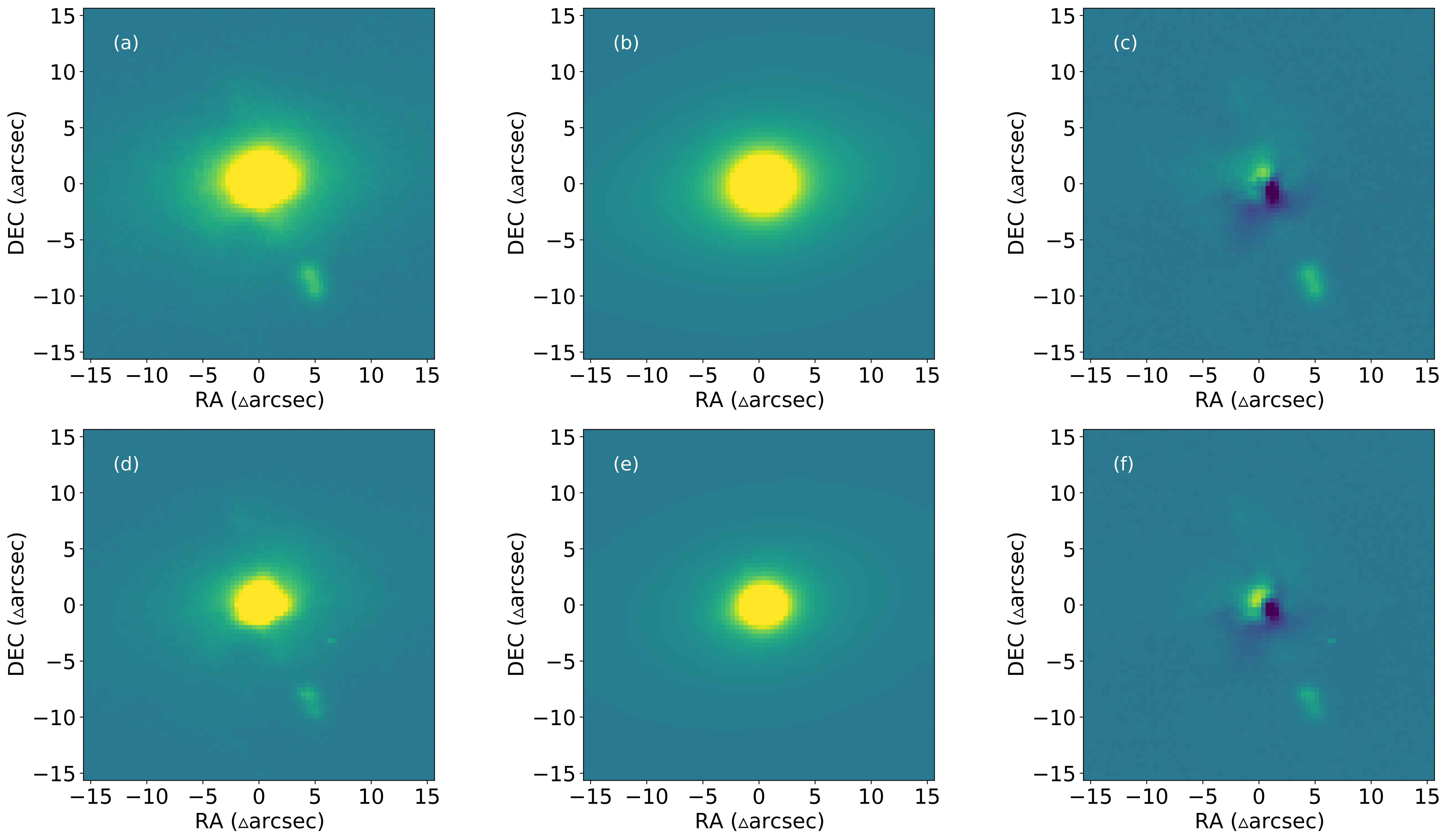}
	\caption{GALFIT fitting results of SDSS r-band (top) and g-band (bottom), original image, GALFIT model image, and residual image (from left to right in each row).}
	\label{fig:GALFIT}
\end{figure*}

Figure~\ref{fig:GALFIT} shows the fitting model and residual image. The residual image tells the goodness of the fitting, which shows some irregular remnants in the central region. Table~\ref{tab:galfit} lists some morphological parameters (e.g., S\'ersic index, axis ratio, position angle of bulge and disk). The parameters derived from g- and r-band are reasonable and consistent with each other. The S\'ersic indexes indicate that PGC 38025 consists of a classical bulge (S\'ersic index = 4.40) and an exponential disk (S\'ersic index = 1.05), consistent with its morphological classification.

\begin{deluxetable*}{c|ccccccc}
    \setlength{\tabcolsep}{4pt}
    \tablewidth{0in}
    \tablecolumns{11}
    \tablecaption{Morphological parameters derived by GALFIT fitting\tablenotemark{} \label{tab:galfit}}
    \tablehead{
        Band & \(n_{b}\) & \(n_{d}\) & \(R b_{\text {axis }}\) & \(R d_{\text {axis }}\) & \(\left(P A_{b}\right)\) & \(\left(P A_{d}\right)\) & \(L_{1} / L_{2}\) 
    }
    \startdata
    	r  & 4.40 $\pm$ 0.16 & 1.05 $\pm$ 0.06 & 0.89 $\pm$ 0.01 & 0.42 $\pm$ 0.02 & -79.39 $\pm$ 3.88 & -81.04 $\pm$ 0.50 & 19.45 - 40.47\\
	    g & 4.21 $\pm$ 0.20 & 0.84 $\pm$ 0.04 & 0.80 $\pm$ 0.01 & 0.50 $\pm$ 0.01 & -80.90 $\pm$ 1.20 & -79.68 $\pm$ 0.39 & 38.81 - 45.41 \\
    \enddata
\end{deluxetable*}

\subsection{CAHA Integral Field Spectroscopy}
\label{sec:CAHA}
\subsubsection{Stellar Population Synthesis (SPS) and Emission Line Measurement}
\label{sec: SPS & EML}
\begin{figure*}
    \centering
    \includegraphics[width=0.75\linewidth]{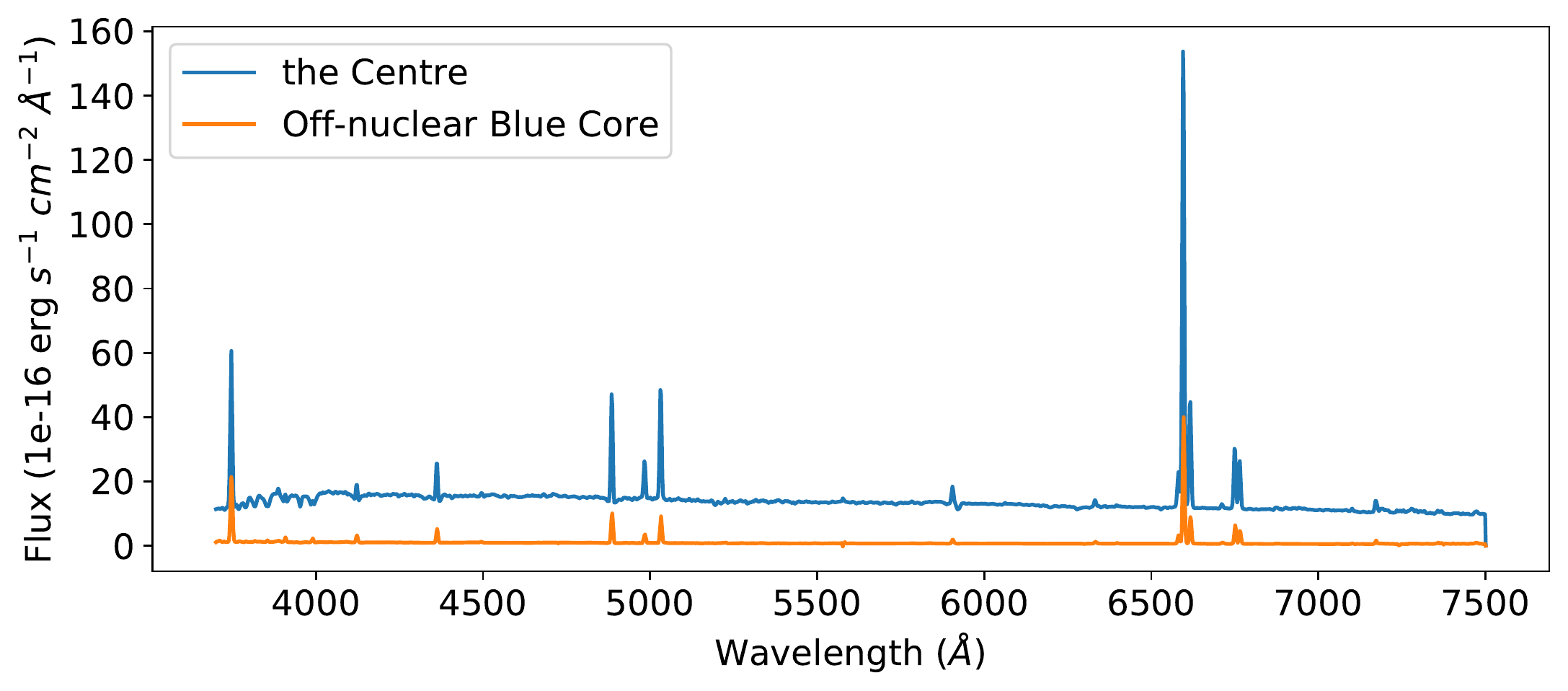}
    \caption{Optical spectra extracted from the galaxy centre (blue) and off-nuclear blue core (orange).}
    \label{fig:spectra-two}
\end{figure*}

As shown in Figure~\ref{fig:spectra-two}, the redshift of off-nuclear blue core is identical to PGC 38025. Both spectra of galaxy centre and the off-nuclear blue core have strong emission-line features, which are indicative of their current star-formation.

We apply the fossil record method of spectral synthesis to recover the 2-dimensional stellar properties. We use the STARLIGHT~\citep{2005MNRAS.358..363C, 2011ascl.soft08006C} to fit  pixel-wise our observed datacube with a set of pre-defined base spectra including all spaxels within a isophote level where the average signal-to-noise ratio is $\geq$ 3. Emission line features are masked during fitting.

The spectral bases are usually made up from evolutionary synthesis models~\citep{2004MNRAS.355..273C, 2005MNRAS.358..363C, 2007MNRAS.375L..16C, 2006MNRAS.370..721M, 2006ApJ...651..861C, 2007MNRAS.381..263A}. Following \citet{2017A&A...608A..27G}, we used a set of 254 single stellar populations (SSPs) by combining GRANADA models of~\citet{2005MNRAS.357..945G} for populations younger than 60 Myr with older SSPs from~\citet{2015MNRAS.449.1177V} based on BaSTi isochrones. Specifically, eight metallicity (Z) were employed (log Z/$Z_\odot$=-2.28, -1.79, -1.26, -0.66, -0.35, -0.06, 0.25, and +0.40), and for each metallicity, stellar age is sampled from 37 templates ranging from 1 Myr to 14 Gyr. A Salpeter Initial Mass Function \citep[IMF,][]{1955ApJ...121..161S} assumption was adopted. We modeled dust effect with the extinction law from~\citet{1989ApJ...345..245C}. 

\begin{figure*}
    \centering
    \includegraphics[width=0.9\linewidth]{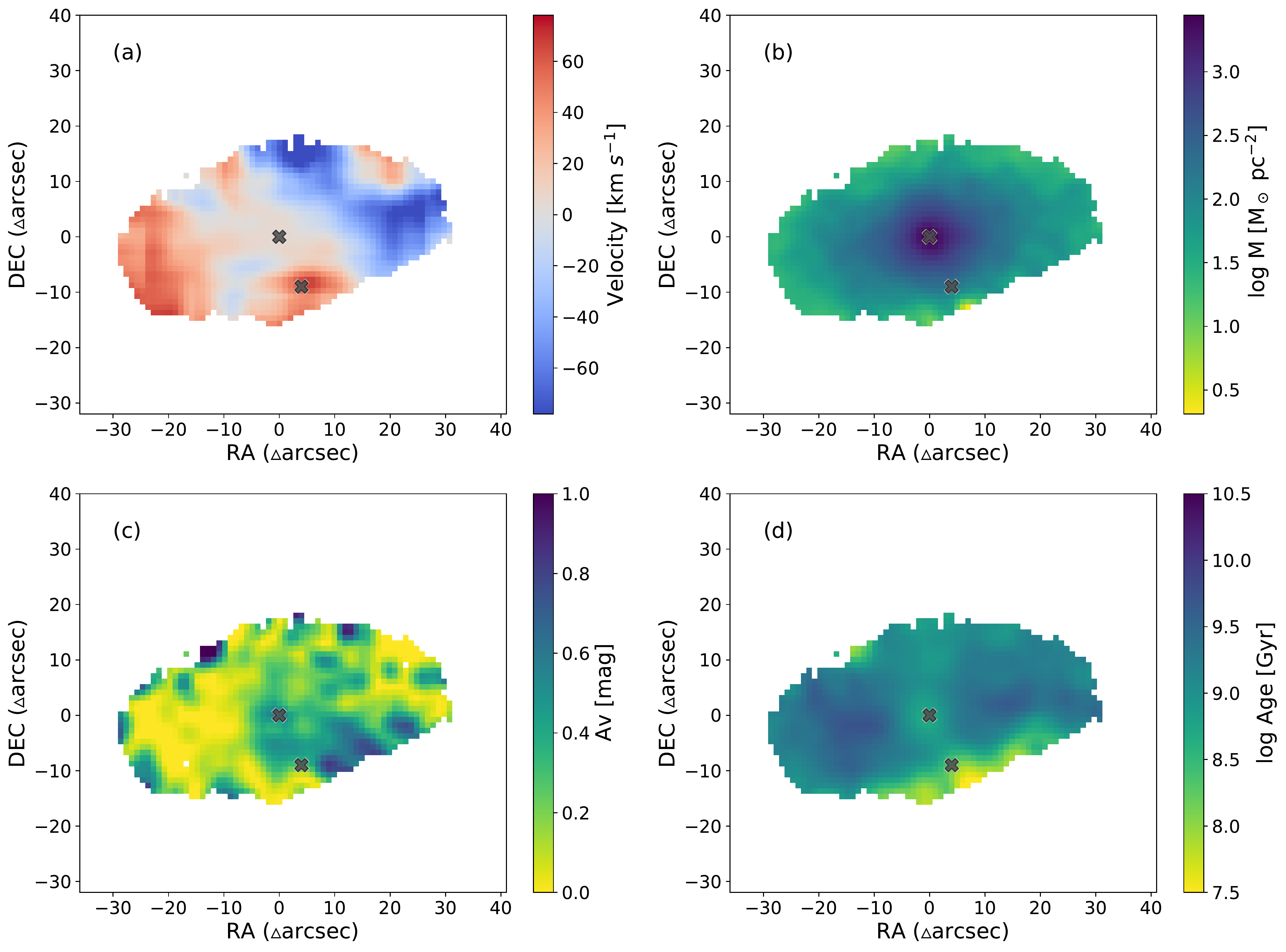}
    \caption{Spaxel-by-spaxel stellar properties distribution of PGC 38025, derived from our stellar population analysis: (a) stellar velocity; (b) stellar mass surface density; (c) dust extinction; and (d) light-weighted age, respectively. The black stars indicate the centres of PGC 38025 and off-nuclear blue core.}
    \label{fig:SPS_4grids}
\end{figure*}

Spaxel-by-spaxel stellar population parameters (e.g, stellar velocity, stellar mass surface density, A$_v$, and light-weighted age) are shown in Figure~\ref{fig:SPS_4grids}. From the stellar velocity map, we can tell that there is a regular rotation generally along the photometric major axis. The off-nuclear blue core located approximately on the photometric minor-axis, and thus supposedly to be the zero-velocity line, shows a redshift velocity of value~$\sim$ + 80 km $s^{-1}$. The stellar mass surface density map, from the centre to the outskirts of the disk, shows a regular decrease. Light-weighted age is slightly younger on the centre of PGC 38025 and off-nuclear blue core. We note that our results would be unaffected with different binning scheme.

We obtained the pure emission-line spectra by subtracting the STARLIGHT fit to the observed spectra, and measured the emission line fluxes (together with their line centres and widths) over the residual spectra using single Gaussian by running the SHerpa IFU line fitting software (\rm SHIFU; Garc\'ia-Benito, in preparation), which was developed based on CIAOs Sherpa package~\citep{2001SPIE.4477...76F, 2007ASPC..376..543D}. 

\subsubsection{Star Formation Rate}
\label{sec: SFR}
\begin{figure*}
    \centering
    \includegraphics[width=0.85
    \linewidth]{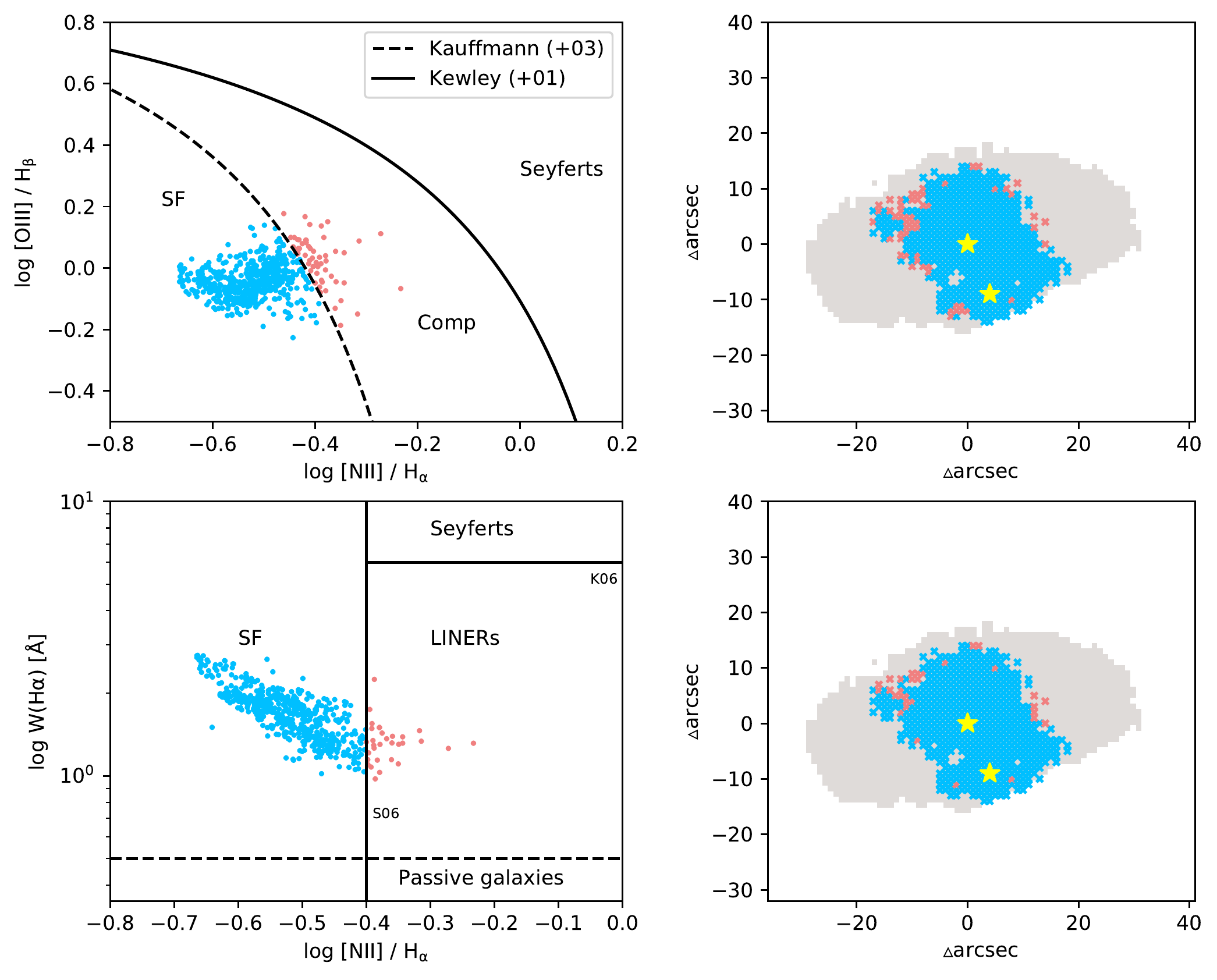}
    \caption{Upper panels are BPT diagram and BPT diagram map. The dashed and solid lines are the boundaries of star-formation, composite and AGN regions from \citet{2001ApJ...556..121K} and \citet{2003MNRAS.346.1055K}, respectively. The blue area represents the SF region, and red area is composite region. Lower panels are WHAN diagram and WHAN diagram map~\citep{2011MNRAS.413.1687C}. The blue, green, and red areas represent SF, LINERs and Seyferts regions, respectively.The line labelled S06 is the \citet{2006MNRAS.371..972S} transposition of the SF/AGN spectral classification scheme, while the line labelled K06 is the transposed \citet{2006MNRAS.372..961K} Seyfert/LINER class division. For diagram maps, the grey background is the region with reliable spectra detection. Yellow stars mark the position of galaxy centre and off-nuclear blue core.}
    \label{fig:BPT}
\end{figure*}
We employed the standard diagnostic diagram [OIII]$\lambda$5007/H$\beta$ versus [NII]$\lambda$6583/H$\alpha$ \citep[hereafter BPT diagram;][]{1981PASP...93....5B} to determine the ionization mechanism of each spaxel. Here we adopted division lines from~\citet{2001ApJ...556..121K} and~\citet{2003MNRAS.346.1055K} to separate the regions into powered by Star Formation (SF), Composite (SF + active galactic nuclei (AGN)), and AGN. A signal-to-noise ratio (S/N) $>$ 5 was applied to all emission lines. The results of BPT diagnostics are shown in upper panels of Figure~\ref{fig:BPT}. Most spaxels belong to SF region, and the rest are composite region defined by~\citet{2003MNRAS.346.1055K}. Alternative, the WHAN diagram \citep{2011MNRAS.413.1687C} was also employed (lower panels of Figure~\ref{fig:BPT}), and shows that PGC 38025 is predominated by SF, which is consistent with BPT diagnostics. We then integrated all spaxels belonging to SF region in WHAN diagram in order to estimate SFR and metallicity for PGC 38025. \citet{Espinosa_Ponce_2020} studied BPT distribution of SF regions, revealing that the location varies with the morphologies of their host galaxies and SF regions in elliptical / lenticular galaxies locate in the centre of [NII]-BPT diagram.

We calculated SFR in bins of regions within deprojected radius (1.0, 1.5, 2.0 effective radius ($R_e$)), generated by using the position angle and axis ratio derived from photometric fitting in Section~\ref{Sec:Morphological Analysis}, and off-nuclear blue core region as well. We adopted the H$\alpha$ (extinction corrected) calibration of SFR from~\citet{1998ApJ...498..541K}, 
\begin{equation}
\rm SFR (M_{\odot}{yr}^{-1})=\frac{L(H_{\alpha, int})}{1.26\times{10^{41}ergs {s}^{-1}}}
\end{equation}

We stacked CAHA IFU spectra for all star-forming regions, and then applied dust extinction correction to them. The extinction was determined by using an IDL package ccm$\_$unred, based on Balmer decrement, assuming a value H$\alpha$/H$\beta$ = 2.86 (case B recombination corresponding to T = ${10}^{4}$ K and electron density $n_e$ = ${10}^2$ ${cm}^{-3}$), and~\citet{1989ApJ...345..245C} extinction law. Strong emission lines, including $H_{\alpha}$ used in SFR calculation, were then fitted with the Gaussian profiles by using an IDL package MPFIT~\citep{2009ATel.2258....1M}. Estimations of SFR for these regions are shown in Table \ref{tab:SFR-Metallicity}, which are 0.446 and 0.023 \rm $M_\odot yr^{-1}$ for PGC 38025 and the blue core region, respectively.

\begin{figure}
    \centering
    \includegraphics[width=1.\linewidth]{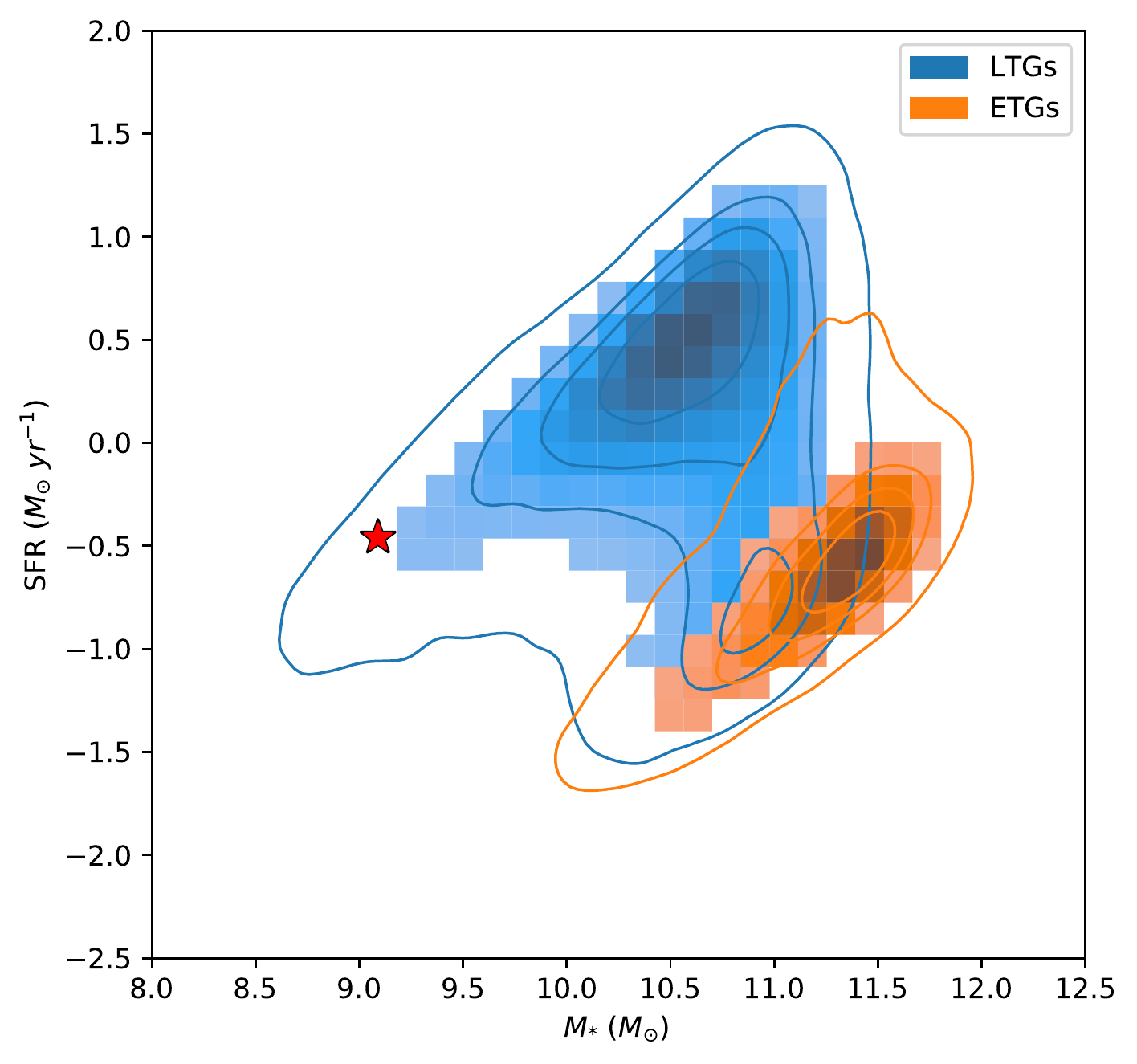}
    \caption{The stellar mass - SFR relation. Blue and orange dots and contours represent distribution of late-type galaxies (LTGs) and early-type galaxies (ETGs) classified in Galaxy Zoo catalog, respectively. The red star represents the value of PGC 38025.}
    \label{fig:SFR-M}
\end{figure}

In Figure~\ref{fig:SFR-M}, we plot the SFR of PGC 38025 within 1.0 $R_e$ on the SFR-${M}_{*}$ relation. We cross match Galaxy Zoo catalog\footnote{\tt Galaxy Zoo is archived at http://zoo1.galaxyzoo.org.} with MPA-JHU~\citep{MPAJHU} DR7 to obtain SFR of early-type galaxies (ETGs; 61,265 galaxies) and late type galaxies (LTGs; 186,745 galaxies), represented as orange and blue dots in Figure~\ref{fig:SFR-M}. The result shows that PGC 38025 follows star-forming main sequence as LTGs do.

\begin{deluxetable}{c|c|c}
	\tablecaption{SFR $\&$ Metallicity} \label{tab:SFR-Metallicity}
	\tablehead{
        Regions & \(\mathrm{SFR}\left(\mathrm{M}_{\odot} \mathrm{yr}^{-1}\right)\) & \(12+\log (\mathrm{O} / \mathrm{H})\)
	}
	\startdata
	    1.0$R_e$  & 0.347 $\pm$ 0.003 & 8.42 $\pm$ 0.01\\
	    1.5$R_e$  & 0.401 $\pm$ 0.002 & 8.42 $\pm$ 0.02\\
	    2.0$R_e$  & 0.446 $\pm$ 0.002 & 8.42 $\pm$ 0.02\\
	    Blue Core  & 0.023 $\pm$ 0.002 & 8.41 $\pm$ 0.01\\
	\enddata
\end{deluxetable}

\subsubsection{Gas-Phase Metallicity}
\label{sec: gas-phase z}
Oxygen abundance is commonly used as a tracer of gas-phase metallicity, as oxygen is the richest ionized species in the optical wavelength range and its production mechanism. The O3N2 index was defined as two ratios of emission lines in~\citet{1979A&A....78..200A},
\begin{equation}
\mathrm{O} 3 \mathrm{N} 2=\log \left(\frac{[\mathrm{O} \mathrm{III}] \lambda 5007}{\mathrm{H} \beta} \times \frac{\mathrm{H} \alpha}{[\mathrm{N} \mathrm{II}] \lambda 6583}\right)
\end{equation}
In contrast to R23\footnote{\tt R23=([OIII]$\lambda\lambda$4959,5007/H$\beta$)/([NII]$\lambda\lambda$6548,84/H$\alpha$)} calibration~\citep{1979MNRAS.189...95P, 2005ApJ...631..231P}, O3N2 has monotonic dependence on the oxygen abundance, and the close distances on wavelength between lines used in both ratios make it merely affected by dust extinction. \citet{2004MNRAS.348L..59P} proposed linear relation between oxygen abundance and line intensity ratio O3N2 (PP04$\_$O3N2 calibration), 
\begin{equation}
   \rm 12 + log(O/H) = 8.74 - 0.31 \times O3N2
\end{equation}
PP04$\_$O3N2 calibration was later superseded by updating calibration proposed in~\citet{Marino_2013} (M13$\_$O3N2) which studied the Te-based abundance of $\sim$600 \rm H\,{\sc ii} regions,
\begin{equation}
    \rm 12 + log(O/H) = 8.533[\pm 0.012] - 0.214[\pm 0.012] \times O3N2
\end{equation}

M13$\_$O3N2 has a wider valid range than PP04$\_$O3N2 (i.e., -1.1 $<$ O3N2 $<$ 1.7 and 12 + log(O/H) $\geq$ 8.0), and is one of the most accurate calibration to date for O3N2, with an intrinsic scatter of 0.18 dex. Our gas-phase metallicity is estimated using M13$\_$O3N2, for the same region as SFR calculation, showed in Table~\ref{tab:SFR-Metallicity}. 
PGC 38025 follow the stellar mass metallicity relation revealed formerly, and there is no obvious difference between PGC 38025 and its off-nuclear \rm H\,{\sc ii} region. 

\subsubsection{Kinematics of ionized Gas}
\label{sec:Kinematics of ionized Gas}
\begin{figure*}
    \centering
    \includegraphics[width=0.95\linewidth]{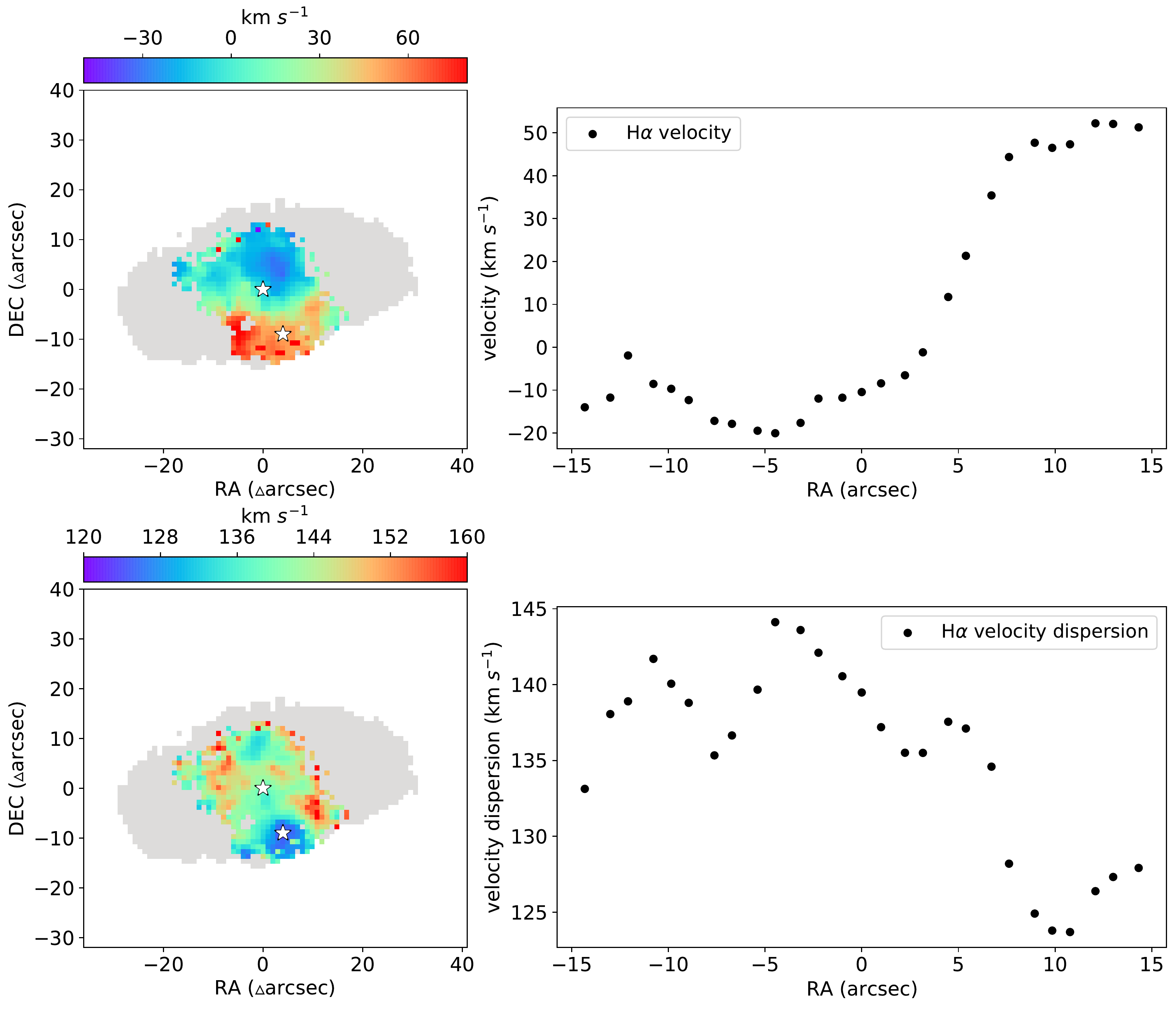}
    \caption{Velocity and velocity dispersion maps of ionized gas in PGC 38025. The white stars indicate the centres of PGC 38025 and off-nuclear blue core, and the red lines show pseudo-slit located at the center of PGC 38025 and tilted across the centre of off-nuclear blue core. Velocity and velocity dispersion along the red line are shown on right two panels.}
    \label{fig:ionized_Velocity}
\end{figure*}
We derived velocity and velocity dispersion maps of ionized gas by fitting $H_\alpha$ emission line, and shown in Figure~\ref{fig:ionized_Velocity}. We extracted the velocity and velocity dispersion along the pseudo-slit located at the center of PGC 38025 and tilted across the centre of off-nuclear blue core. Velocity along the pseudo-slit is asymmetric and varies vary from - 20 to + 50 km $s^{-1}$. The off-nuclear blue core located at 9\farcs{85} away from the centre, where velocity is at the peak value of \rm$\sim$ 50 km $s^{-1}$, while the velocity dispersion reaches the bottom value of \rm$\sim$ 125 km $s^{-1}$. We investigated the stellar velocity along this pseudo-slit, and found it follows the similar trend as ionized gas. We note that there is a clump north-east to the off-nuclear blue core, who reaches the spatial local velocity peak value of \rm$\sim$ + 100 km $s^{-1}$. 

\subsection{NOEMA Data Analysis}
\label{sec: NOEMA data analysis}
\subsubsection{CO (1-0) Spectrum and Moments Analysis}
\label{sec: CO spectrum and moment analysis}
\begin{figure*}
    \centering
    \includegraphics[width=0.9\linewidth]{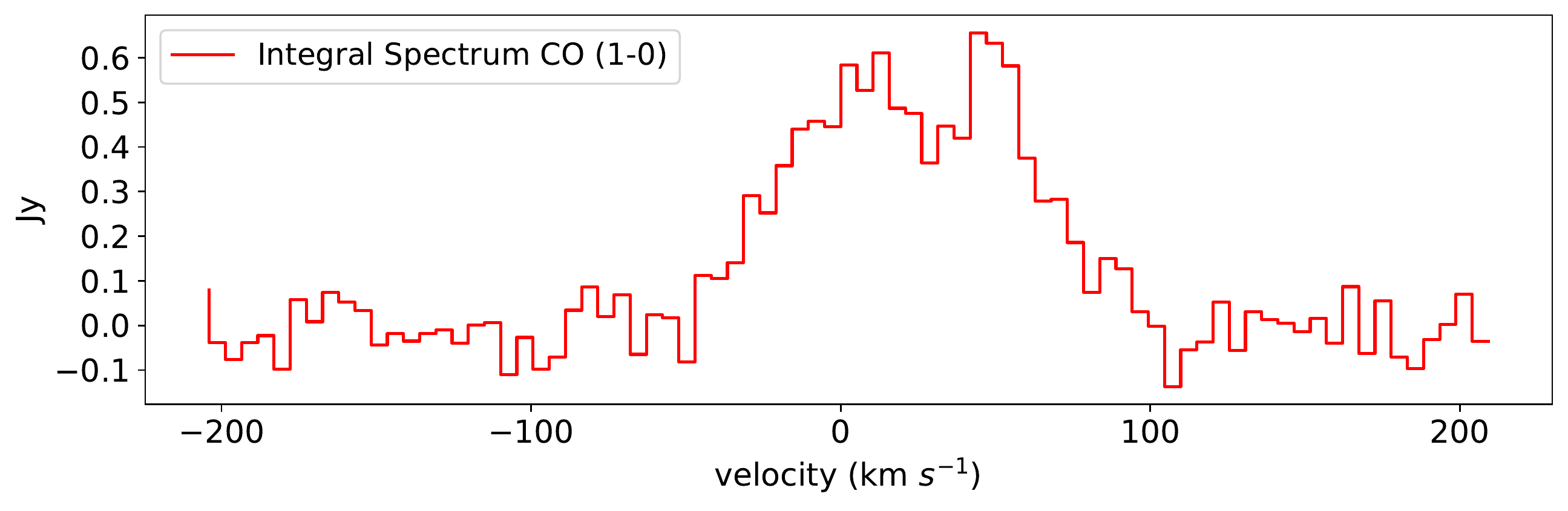}
    \caption{Integral Spectrum of CO J=1$\rightarrow$0 in PGC 38025.}
    \label{fig:CO-spectrum}
\end{figure*}
\begin{figure*}
    \centering
    \includegraphics[width=1.0\linewidth]{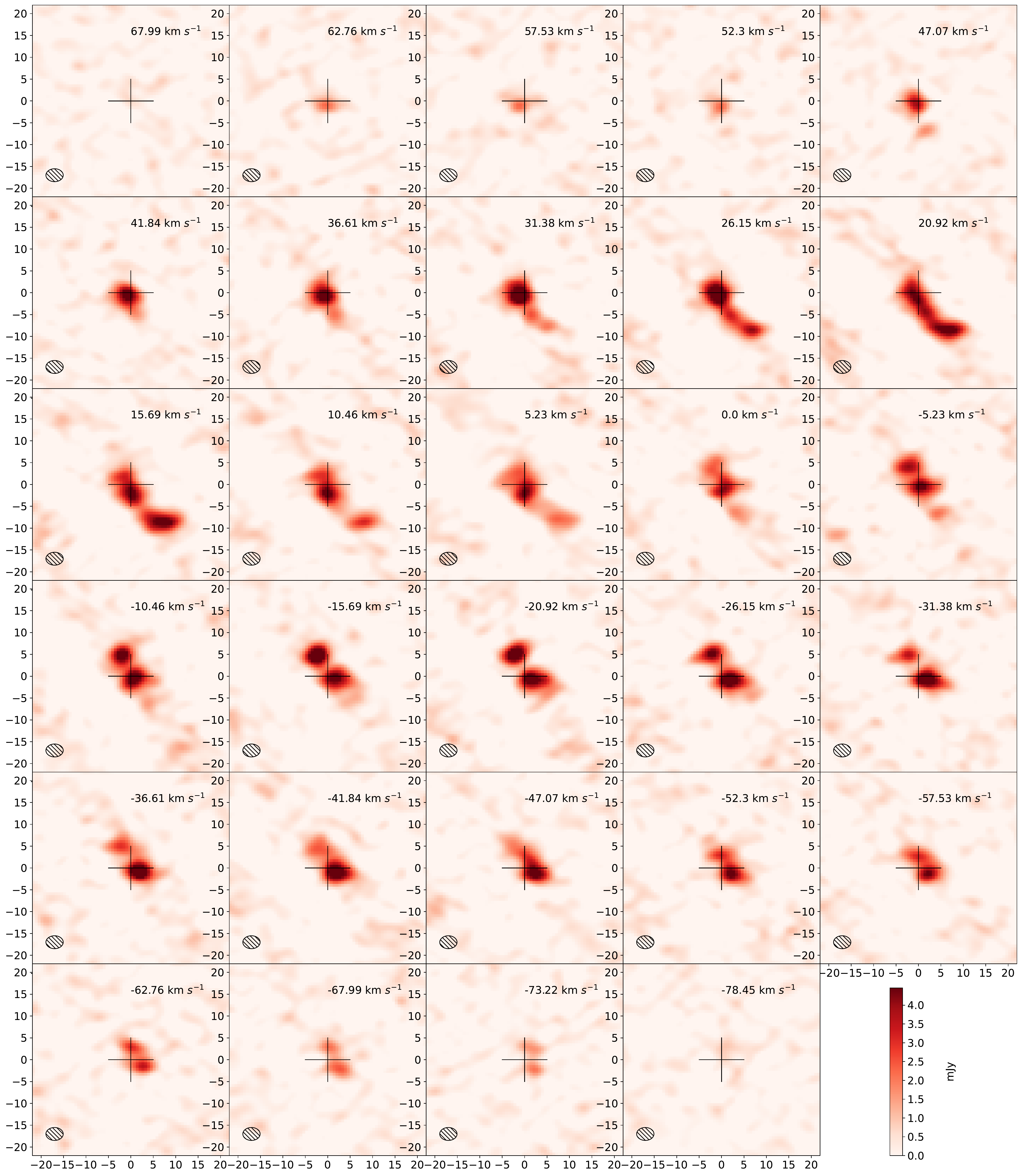}
    \caption{Channel by channel images of CO J=1$\rightarrow$0 observation centred on it’s rest frame frequency. In each panel, the beam size of observation is shown at left bottom, the relative velocity is shown at upper right, the black cross denote the centre of galaxy PGC 38025. Scale is the same for all panels and the color bar is shown at the right bottom.}
    \label{fig:CAHA-optical+CO}
\end{figure*}

CO J=1$\rightarrow$0 is clearly detected for PGC 38025, and the integrated spectrum shows double-horn feature, as shown in Figure~\ref{fig:CO-spectrum}. This spectrum profile cannot be well fitted with two Gaussians, indicating that there are multiple molecular components. To explore these components, we show the channel images of CO J=1$\rightarrow$0 in Figure~\ref{fig:CAHA-optical+CO}. We corrected the centre of the channel image to be consistent with the optical centre of PGC 38025 and mark the centre with black cross. The velocity value shown on the upper right corner in each panel has been corrected to the rest frame. These channel maps show that there are molecular clumps appear at the optical centre of PGC 38025 and location of off-nuclear blue core. There are also clumps showing north east to the optical centre of PGC 38025, mainly appear with negative velocity values (regressing along the light-of-sight). 

\begin{figure*}
    \centering
    \includegraphics[width=0.9\linewidth]{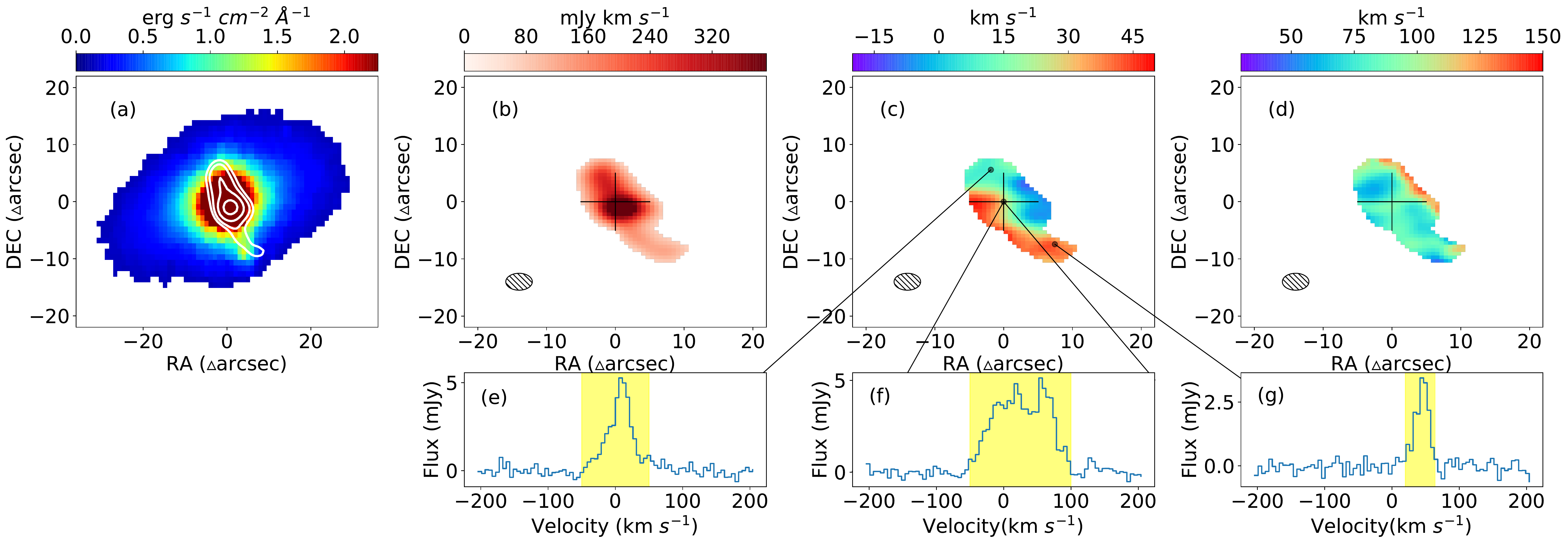}
    \caption{(a) The background is synthesis r-band image of CAHA observation, contour represents the distribution of CO (1-0) emission; (b) flux map (moment 0) of CO (1-0); (c) velocity map (moment 1) of CO (1-0); (d) velocity dispersion map (moment 2) of CO (1-0); (e) spectrum of point symmetric to off-nuclear blue core; (f) spectrum of centre of PGC 38025; (g) spectrum of off-nuclear blue core.}
    \label{fig:CO_Moment}
\end{figure*}

We generated the moments maps (from left to right are flux, velocity and velocity dispersion), and shown in right panels of Figure~\ref{fig:CO_Moment}. We plot the spectra for centre of PGC 38025, the off-nuclear blue core, and the point symmetric to the off-nuclear blue core as well on the right bottom panels. To demonstrate the spatial distribution of CO detection, we plot the flux contour of CO (1-0) on the background of CAHA synthesis SDSS-r band flux (left panel of Figure~\ref{fig:CO_Moment}). The molecular gas distributed along the photometric minor-axis is consistent with the distribution of the ionized gas.

\begin{figure*}
    \centering
    \includegraphics[width=0.8\linewidth]{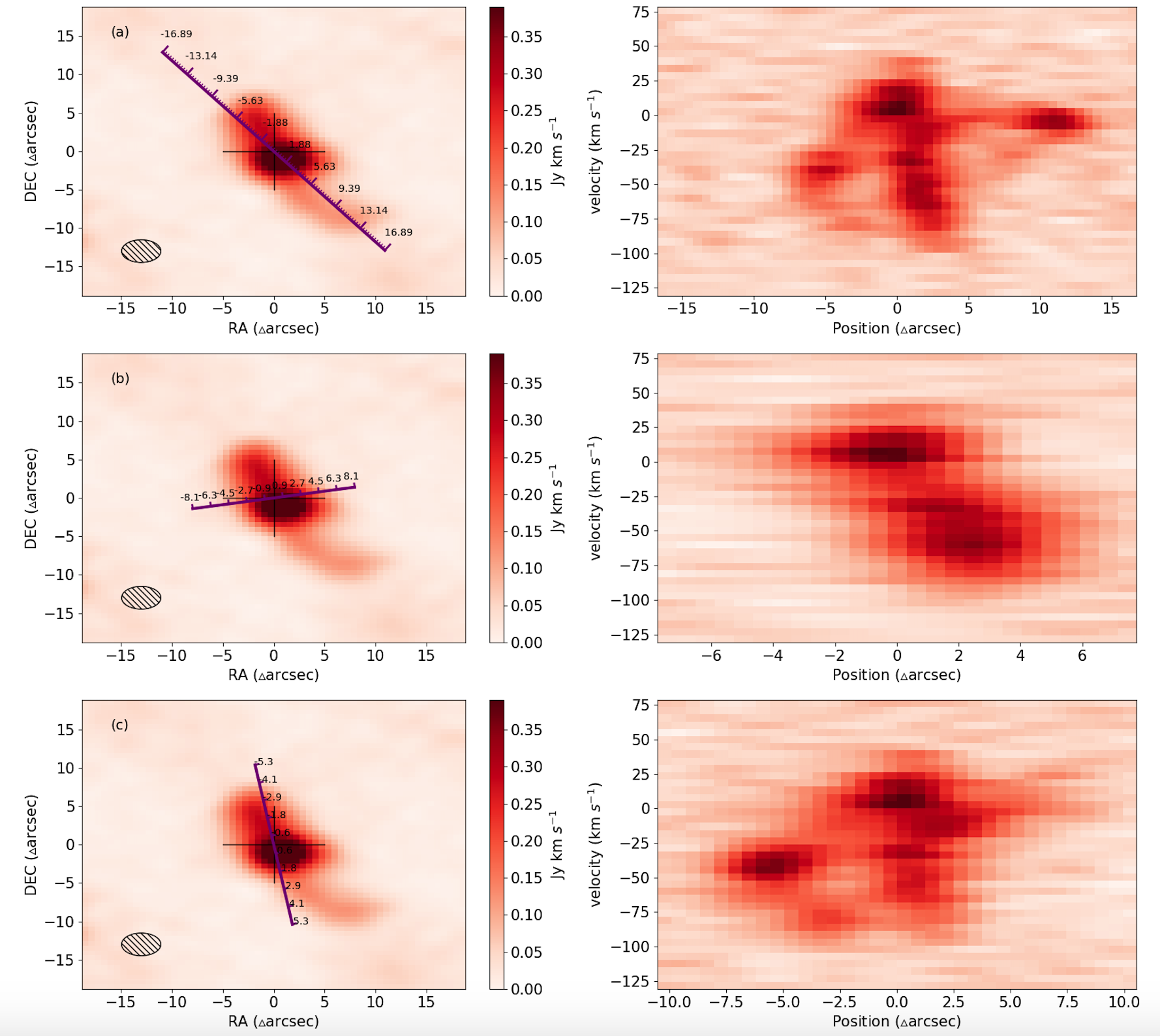}
    \caption{Left panels show the CO (1-0) flux of PGC 38025 overlapped pseudo-slits used to extract the position velocity map, which shown in the right panels. Slit description: (a) slit across the galaxy centre and off-nuclear; (b) slit along the galaxy photometric major axis; (c) slit along the galaxy photometric minor axis. Slit orientation and position are shown by the purple scale ruler in left panels.}
    \label{fig:PV-map}
\end{figure*}

To inspect the kinematics of molecular components, we take a 1''-width pseudo-slit locating in the optical centre of PGC 38025 and tilted to across the off-nuclear blue core, and extract the Position-Velocity (PV) map from cleaned lmv spectra cube, as shown in first row of  Figure~\ref{fig:PV-map}. The complex velocity structure, which is obviously not a single rotating disk, consists of a suspicious rotation along the photometric minor-axis. Together with the molecular velocity map, it also shows another independent rotation in the narrower nuclear region along the major axis, aligned with the stellar components. Another two PV maps are extracted along pseudo slits along the orientations of photometric major and minor axis of PGC38025, as shown on the second and third rows of Figure~\ref{fig:PV-map}. PV map along the photometric major axis present a rotation disk, though not well resolved, while the one along photometric minor axis is messy.

\begin{figure}
    \centering
    \includegraphics[width = 0.9\linewidth]{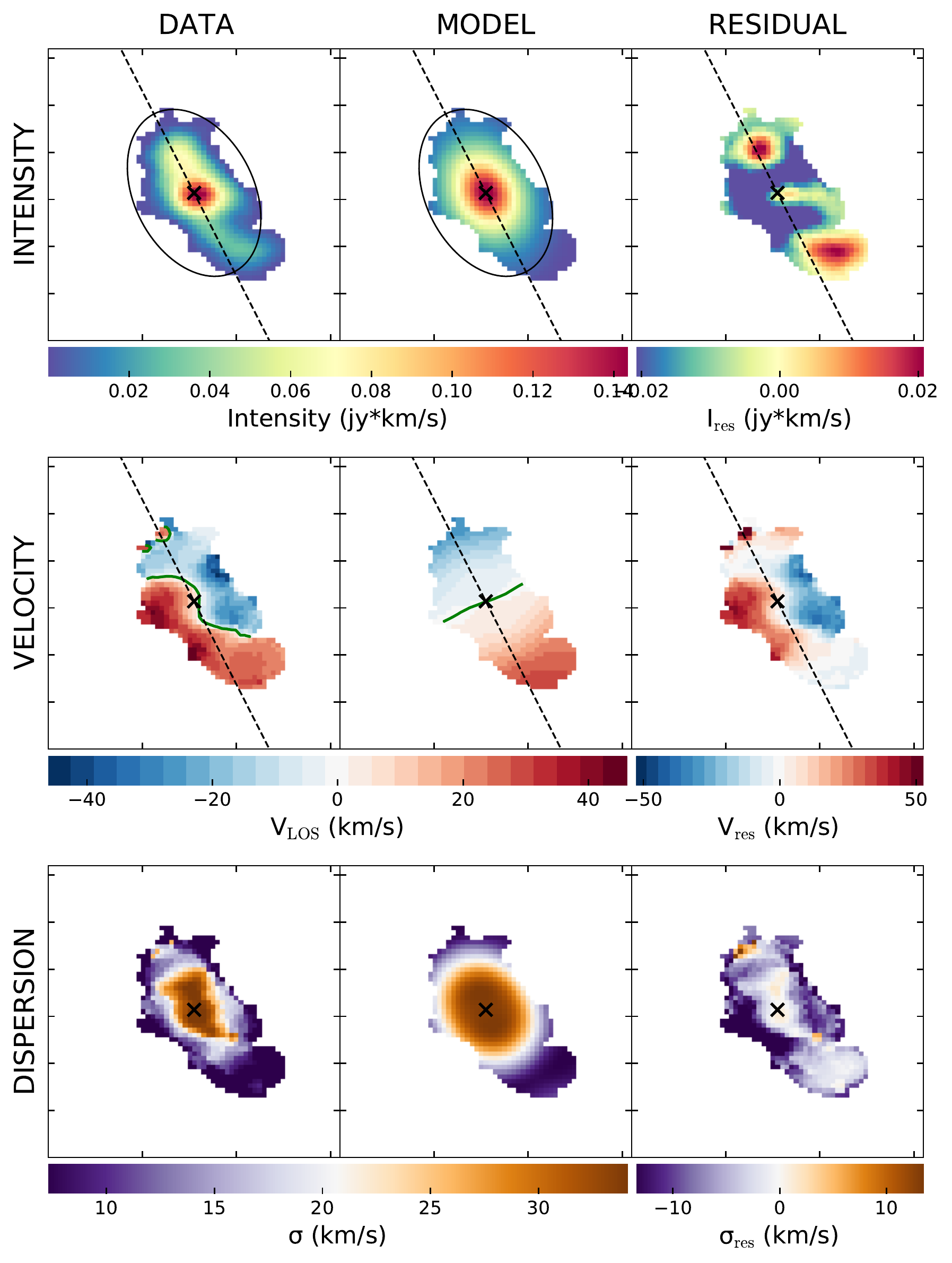}
    \caption{Moments maps of $^{\rm 3D}$Barolo modelling. From the upper to bottom are intensity, velocity, and velocity dispersion maps. Each row from left to right are origin, model, and residual maps.}
    \label{fig:Bbarolo_moments}
\end{figure}

\begin{figure}
    \centering
    \includegraphics[width = 0.9\linewidth]{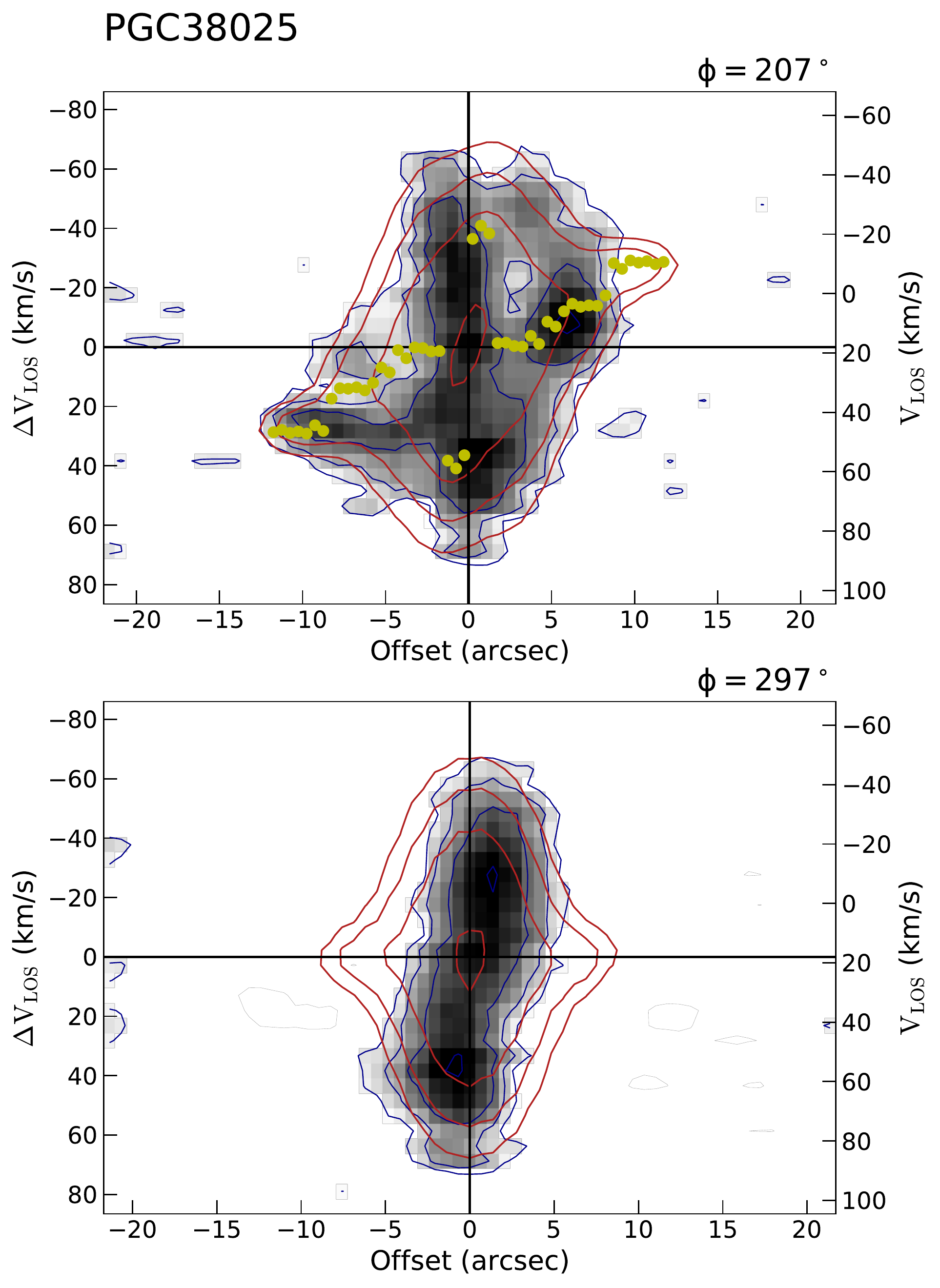}
    \caption{PV maps of $^{\rm 3D}$Barolo modelling, alonging the major and minor axis. Using CO data, is shown in grey and blue contour. The model is shown with red contours, and yellow dots shows the resulting rotation curve of the galaxy, projected on the data.}
    \label{fig:Bbarolo_pv}
\end{figure}

We also employed the  $^{\rm 3D}$Barolo software~\citep{2015MNRAS.451.3021D} to model the kinematics of molecular gas, resulting in consistent results. The moments maps are shown in Figure~\ref{fig:Bbarolo_moments}, and the PV maps are shown in Figure~\ref{fig:Bbarolo_pv}.


\subsubsection{Molecular Hydrogen Mass Estimation}
\label{sec: molecular hydrogen mass estimation}
CO line luminosity was calculated using equation from~\citet{2005ARA&A..43..677S},
\begin{equation}
L_{\mathrm{CO}}^{\prime}=3.25 \times 10^{7} S_{\mathrm{CO}} \Delta v \nu_{\mathrm{obs}}^{-2} D_{\mathrm{L}}^{2}(1+z)^{-3}
\end{equation}
\noindent
where $L_{\mathrm{CO}}$ is luminosity of CO J=1$\rightarrow$0 line luminosity measured in K km $s^{-1}$, \rm$S_{\mathrm{CO}} \Delta v$ is the velocity integrated flux measured in Jy km \rm$s^{-1}$, \rm$\nu_{\mathrm{obs}}$ is the observed frequency in GHz, and the luminosity distance $D_{\mathrm{L}}$ in Mpc.

We then adopted the molecular gas mass to CO luminosity ratio of the Galaxy~\citep{1991IAUS..146..235S}, 
\begin{equation}
\alpha \equiv M_{\mathrm{gas}} / L_{\mathrm{CO}}^{\prime}=4.6 \mathrm{M}_{\odot}\left(\mathrm{K} \mathrm{km} \mathrm{s}^{-1} \mathrm{pc}^{2}\right)^{-1}
\end{equation}
\noindent
The molecular hydrogen mass derived is \rm$log (H_2) = 8.07$. \citet{2013ApJ...769...82S} estimated the mass of molecular hydrogen, from observation of both CO J=1$\rightarrow$0 and CO J=2$\rightarrow$1 using IRAM 30m telescope, and obtained the value of 8.29 $\pm$ 0.01. The molecular hydrogen mass fraction ($\rm M_{H_2}$ / $\rm M_\star$) is 9.47 $\rm \%$. 

We estimate the star formation efficiency (SFE; defined as SFR per unit molecular gas mass, SFR/$M_{H_{2}}$), using our molecular gas mass and SFR derived, obtained the value of 2.2 $\times$ $\rm 10^{-9}$ $\rm yr^{-1}$. Assuming a constant consumption of gas, the depletion time is 0.45 Gyr. According to literature, the spiral galaxies of \citet{1998ApJ...498..541K} have average SFEs of $\approx$ 1.5 $\times$ $10^{-9}$ $\rm yr^{-1}$, and the medium of SFEs for $\rm {ATLAS}^{3D}$ early type galaxies is $\approx$ 4 $\times$ $10^{-10}$ $yr^{-1}$ \citep{2015MNRAS.449.3503D}. The molecular gas consumption in PGC 38025 is characterized closer to late-type star-forming galaxies.

\section{Discussion}
\label{sec: discussion}
\subsection{Kinematics: Polar Disc Structure? External Originated Gas?}
\label{sec: discussion1}
Gas, in all phases, is one of the most important factor regulating galaxy evolution through both internal and external processes. Galaxies can internally recycle gas through stellar mass loss. On the other hand, external gas supply includes (minor and major) merger, cooling flow from halo, and gas accretion. Moreover, external processes are observationally supported to be the main pathway for early-type galaxies in the field environment \citep{1989ApJS...70..329K, 2006MNRAS.366.1151S, 2006MNRAS.373..906M, 2010MNRAS.409..500O, 2011MNRAS.417..882D, 2012ApJ...748..123S, 2015MNRAS.446..120D}.

Kinematical decoupling between gaseous and stellar components is found in early-type galaxies \citep{2006MNRAS.366.1151S, 2011MNRAS.417..882D, 2014MNRAS.444.3388S}. According to the angular momentum conservation, gas produced by internal process (e.g., stellar mass loss) will be aligned to the stellar kinematics, and thus it is commonly accepted that kinematically misaligned gas originates externally. Moreover, ionized, molecular and atomic gas in early-type galaxies are found always aligned with each other, even though may be misaligned with stellar components \citep{2011MNRAS.417..882D, 2016MNRAS.463..913J, 2019arXiv191204522L}. Simulations show that both the episodic and continuous gas accretion \citep{1996ApJ...461...55T, 1998ApJ...506...93T} and merging with companions \citep{2011MNRAS.416.1654B, 2014MNRAS.444.3357N} can create the kinematically misaligned components in the host galaxies.

We obtained kinematics of stellar components, ionized and molecular gas from CAHA optical and NOEMA CO J=1$\rightarrow$0 observation for PGC 38025. We found that the stellar disk rotates along the photometric major-axis, while both ionized and molecular gas distribute and rotate approximately along the minor-axis. These orthogonal rotational pattern supports that the gas in PGC 38025 is originated externally. However, accretion from cosmic web, which usually generates a metal-poor gas reservoir, can be ruled out. Considering that the gas-phase metallicity of off-nuclear blue core is nearly the same as global value of PGC 38025, we propose the scenario that PGC 38025 result from gas-rich minor merger, i.e., the gas externally originated with initial angular momentum different from stellar component in PGC 38025. Co- or counter-rotation, rather than orthogonal one in PGC 38025, are the most stable configurations of gas. Simulation in \citet{2012MNRAS.425.1967S} shows that merger following with continuous gas accretion can create and maintain this polar disc structure for several Gyr.

From the PV map as shown in Figure~\ref{fig:PV-map}, we found that beside the rotation along the photometric minor-axis (misaligned with stellar components) with radius of $\sim$ 10$\farcs$, there is another rotation along the major-axis in a narrower region (with radius $\sim$ 5$\farcs$). We found a simulation in \citet{2015MNRAS.451.3269V} that may explain this structure. The misaligned externally originated gas realigns with the stellar components when the torque generated by stellar components dominates over the gas. During the realigning process, the central region aligns faster than the outskirt and generates a warped gaseous disk. The warped structure of molecular gas disk is indicated by the faint connections on the PV map between clumps, though it cannot be well resolved with our current spatial resolution of the data. This differential realignment of polar disk is reasonable in consideration of the central deeper potential well generated by the dense bulge (with S\'ersic index = 4) hosted by PGC 38025.

\subsection{Nature of off-nuclear blue core? What triggers star formation in S0 PGC 38025?}
PGC 38025 is a field S0 galaxy, whose sparse environment leads to the star formation possibility of gas obtained externally from minor merger or gas accretion, supported by gas-stellar misalignment. As mentioned above, the redshift of off-nuclear blue core is identical to PGC 38025, and the spectrum of the off-nuclear blue core has strong emission-line feature and nearly zero continuum. Through BPT diagnostics, this off-nuclear blue core was determined to be an \rm H\,{\sc ii} region excited by star-formation. CO detection further revealed the molecular gas served as reservoir (gas fraction $M_{H_2}$ /$M_*$ reaches $\sim$ 10$\%$) of PGC 38025, supplying for star formation on an extended region, including this blue core. The external gas (ionized and molecular) built a polar disk in PGC 38025. We suggest this blue core as a part of this disk, relative denser and thermal cold (with low velocity dispersion), generated during misaligned gas accretion and rotation. The SFR of the blue core is 0.023 $M_{\odot}yr^{-1}$. Gas-phase metallicity of the blue core is 8.41, nearly the same as the global value, which is reasonable as the global value might be elevated by former chemical enrichment.

A merger would precisely tend to flatten the metallicity gradient. We checked the gas-phase metallicity radial distribution by divide the galaxy into 5 ring regions, each with width of 0.4 Re. the metallicity gradient is flat withing the uncertainties which is consistent with the merger or accretion scenario we proposed.

\subsection{Rejuvenation of lenticular galaxies? }

Recent star formation rarely happened in lenticular galaxies. In rejuvenated lenticular galaxies, star formation activity is usually centralized, e.g., \citet{2010ApJ...714L.171T} and \citet{2020ApJ...889..132G}. It is a reasonable consequence of star formation provoked after the gas-inflow following the gravitational well of a galaxy. However, PGC 38025 is a rejuvenating S0, and its star formation is spatially extended (Fig.~\ref{fig:BPT}).

Our study, taking advantage of two aspects, makes an important progress on advancing recognition of lenticular evolution. On the one hand, we applied both spatial resolved optical spectroscopic and radio interferometric observations, and obtained more complete information than photometric, single slit spectroscopic or single aperture CO detection. On the other hand, based on these observations, PGC 38025 shows both central and off-nuclear star-formation activities before external originated gas relaxed (i.e., the angular momentum of gas is still perpendicular to the angular momentum of former exist stellar components). Through this pilot case study, combined IFS and radio interferometric observation, we call for attention on lenticulars with recent star-formation (Detailed discussion of kinematics of stellar, ionized, and molecular gas is in Section~\ref{sec: discussion1}. Refer to Fig~\ref{fig:SPS_4grids}, \ref{fig:ionized_Velocity}, \ref{fig:CO_Moment}, \ref{fig:PV-map}). 

\citet{10.3390/galaxies4010001} and \citet{2016MNRAS.463..913J} found the existence of polar gaseous disks and/or stellar-gas misalignment in star-forming lenticular galaxies, which is exactly happening in PGC 38025 as we found. For the cold gas aspect, due to the lack of molecular gas in lenticulars, most studies are based on atomic hydrogen observation \citep{1991A&A...243...71V}, which is not a direct star-formation tracer. Limited CO detection of lenticulars are usually based on single aperture \citep{2002PASJ...54..555K, 2003ApJ...584..260W}. Resolved optical data (i.e., IFS) shows the kinematics of stellar and ionized gas, corresponding to the accumulative and recent star-formation. Neither with only spatially resolved optical observation nor combining with unresolved CO detection can reveal the complete scenario of rejuvenation in lenticulars. In this work, we are able to investigate the connection between molecular gas and star formation by combining spatially resolved molecular observation.

\section{Summary}
\label{sec: summary}
PGC 38025 is a star forming early type galaxy, which morphologically holds a classical bulge and an exponential disk. The off-nuclear blue core exhibited in PGC 38025 is identified as an \rm H\,{\sc ii} region through its spectral feature and BPT diagnostics. Properties of PGC 38025 are listed in Table~\ref{tab:properties_pgc38025}.

 Star formation rate of PGC 38025 and its accompanying blue core are 0.446 and 0.023 $\rm M_{\odot}$ $\rm yr^{-1}$, respectively. The star formation rate is relatively high regarding its stellar mass, while following the general stellar mass-SFR relation of former work. The metallicity 12+log(O/H) of them are 8.42 and 8.41, respectively. The metallicity of off-nuclear core is the same within the errors, which indicates that this blue core may be the same origin with the galaxy PGC 38025, or their ISM is already well-mixed during galaxy evolution.
 
The kinematics derived from CAHA's optical data and IRAM-NOEMA's millimeter data together support that this source owns a rather complex kinemetry for ionized gas, molecular gas, and stellar components. There is a stellar disk rotating along the major axis, and an ionized gas disk rotating along the minor axis. These two present older stellar populations within galaxy PGC 38025 and newborn stars, respectively. As for molecular gas, besides the rotation along the photometric minor-axis, there is a nuclear rotation along photometric major-axis as well. All the kinematics of different components in PGC 38025 support it to be formed from a gas-poor progenitor obtained gas from a wet-merger/gas accretion, generating an orthogonal gaseous disk and also a blue knot (the off-nuclear blue core) on the polar gaseous disk.

Limited by the angular resolution and the interferometric filtering effect, the kinematic structure is not well-resolved in the central nuclear region, and the large scale structures are still missing. We also find that there is another high velocity component decreasing from the centre to the blue core (see the PV diagram), which may also be an indication of bipolar outflows. 

We will further explore the case in a future work in the light of granted VLA HI data.

\begin{deluxetable}{cccc}
	\tablecaption{Properties of PGC 38025 \tablenotemark{} \label{tab:properties_pgc38025}}
	\tablehead{
        Quantity & Value & Unit & Ref.
	}
	\startdata
        Hubble Type & S0 & \(-\) & \((1,2)\) \\
        RA & \(12: 02: 37.195\) & \(-\) & \((1)\) \\
        Dec & \(64: 22: 29.070\) & \(-\) & \((1)\) \\
        Redshift & \(0.00505\) & \(-\) & \((1)\) \\
        Distance & \(21.71\) & \(\mathrm{Mpc}\) & \((2)\) \\
        \(\mathrm{M}_{\text {stellar }}\) & \(1.23 \times 10^{9}\) & \(\mathrm{M}_{\odot}\) & \((3)\) \\
        \(\mathrm{M}_{\mathrm{H}_{2}}\) & \(1.17 \times 10^{8}\) & \(M_{\odot}\) & \((2)\) \\
        molecular gas fraction & \(9.47 \%\) & \(-\) & \((2)\) \\
        SFR & \(0.46\) & \(\mathrm{M}_{\odot} y r^{-1}\) & \((2)\) \\
        \(12+\log (\mathrm{O} / \mathrm{H})\) & \(8.42\) & \(-\) & \((2)\) \\
	\enddata
	\tablecomments{Reference: (1) NED, (2) this paper, (3) MPA-JHU}
\end{deluxetable}

\software{PyCASSO~\citep{2017MNRAS.471.3727D}, SHIFU (Garc\'ia-Benito, in preparation), GILDAS~\citep{2013ascl.soft05010G}, STARLIGHT~\citep{2005MNRAS.358..363C, 2011ascl.soft08006C}, GALFIT \citep[version 3.0.5;][]{2002AJ....124..266P, 2010AJ....139.2097P}, $^{\rm 3D}$Barolo~\citep{2015MNRAS.451.3021D}, numpy~\citep{5725236}, Matplotlib~\citep{2007CSE.....9...90H}, pvextractor~\citep{2016ascl.soft08010G}, astropy~\citep{astropy:2013, astropy:2018}}

\section*{Acknowledgements}
This work is supported by the National Key Research and Development Program of China (No. 2017YFA0402703) and by the National Natural Science Foundation of China (No. 11733002). This work is based on observations carried out with the IRAM NOrthern Extended Millimeter Array. IRAM is supported by INSU/CNRS (France), MPG (Germany) and IGN (Spain). In addition, we acknowledge the supports of the staffs from CAHA and NOEMA, especially Orsolya Feh\'er. R.G.B. acknowledges financial support from theState Agency for Research of the Spanish MCIU throughthe “Center of Excellence Severo Ochoa” award to the Instituto de Astrof\'isica de Andaluc\'ia (SEV-2017-0709) and grants PID2019-109067GB-I00 and P18-FRJ-2595.

\bibliography{PGC38025}{}
\bibliographystyle{aasjournal}

\end{document}